\documentclass[12pt,reqno]{amsart}
\usepackage{mathrsfs}
\usepackage{amsmath}
\usepackage{amsxtra}
\usepackage{amscd}
\usepackage{amsthm}
\usepackage{amsfonts}
\usepackage{amssymb}
\usepackage{pstricks}
\usepackage{pstricks,pst-node}

\newcommand{\mysection}[1]{\section{#1}\setcounter{equation}{0}}

\newtheorem{theorem}{Theorem}[section]

\theoremstyle{definition}
\newtheorem{definition}[theorem]{Definition}

\theoremstyle{remark}

\numberwithin{equation}{section}

\def\prf{{\textit {Proof:} }}

\newcommand{\R}{{\mathbb R}}

\newcommand{\Eq}{{\mathcal Eq}}

\newcommand*{\f}[2]{\frac{#1}{#2}}

\newcommand*{\csct}{\csc{\theta}}
\newcommand*{\sint}{\sin{\theta}}
\newcommand*{\cost}{\cos{\theta}}

\newcommand*{\pr}[1]{\frac{\partial #1}{\partial r}}
\newcommand*{\pu}[1]{\frac{\partial #1}{\partial u}}
\newcommand*{\pth}[1]{\frac{\partial #1}{\partial \theta}}
\newcommand*{\pf}[1]{\frac{\partial #1}{\partial \phi}}

\newcommand{\ls}{\setlength{\baselineskip}{12pt}
                 \setlength{\parskip}{3mm}}

\begin{document}


\title[Peeling property]{Peeling property of Bondi-Sachs metrics for nonzero cosmological constant}

\author[]{Fangquan Xie, Xiao Zhang}
\address[]{Institute of Mathematics,
Academy of Mathematics and Systems Science, Chinese Academy of Sciences, Beijing 100190, China}
\address[]{School of Mathematical Sciences, University of Chinese Academy of Sciences, Beijing 100049, China}
\email{fangquanxie@amss.ac.cn, xzhang@amss.ac.cn}

\begin{abstract}
In this paper, we show that the peeling property still holds for Bondi-Sachs metrics with nonzero cosmological constant under the boundary condition given by Sommerfeld's radiation condition together with three nontrivial $\Lambda$-independent functions $B$, $a$, $b$. This should indicate the new boundary condition is natural. Moreover, we construct some nonstationary vacuum Bondi-Sachs metrics without Bondi news, which Newmann-Penrose quantities fall faster than usual. This provides a new feature of gravitational waves for nonzero cosmological constant.
\end{abstract}

\keywords{Bondi-Sachs metric, peeling property, cosmological constant}

\maketitle

%
%
\mysection{Introduction}\ls
The full nonlinear theory of gravitational waves was established in the work of Bondi, Sachs, et al. in framework of Bondi-Sachs metrics \cite{Bondi, Sachs}. Lately, the theory was re-formulated by Penrose in a more geometric notion of conformal compactification of spacetimes \cite{Penr}. However, they were studied well only when the cosmological constant is zero. In this case, by assuming Sommerfeld's radiation condition, there is a natural boundary condition to make the vacuum Bondi-Sachs metrics asymptotic to
\begin{eqnarray*}
 \begin{aligned}
g =&-\Big(1-\frac{2M}{r}\Big)du ^2 -2du dr+2\Big(c _{, \theta} +2c \cot \theta +d _{, \psi} \csc \theta \Big) dud\theta \\
   &+2 \Big(d_{, \theta} +2d \cot \theta -c _{,\phi} \csc \theta\Big)\sin \theta du d\phi \\
   &+r ^2 \Big[(1+\frac{2c}{r})d\theta ^2 +\frac{4d}{r}\sin \theta d
     \theta d \phi+(1-\frac{2c}{r})\sin ^2 \theta d \phi ^2\Big]
 \end{aligned}
\end{eqnarray*}
where $u$ is retarded coordinate, $r$, $\theta $ and $\phi$ are polar coordinates,
$M$, $c$, $d$ are smooth functions of $u$, $\theta$, $\phi$ defined on $\R \times S ^2$ with regularity condition
$\int _0 ^{2\pi} c(u, \theta, \phi)d\phi =0$ for $\theta =0, \pi$ and for all $u$.

The functions $c_{,u}$, $d_{,u}$ are {\em Bondi news} which characterizes the presence of gravitational waves. Geometrically, they measure the deformation of standard 2-sphere as a result of gravitational waves.
The Bondi energy-momentum defined on $u$-slice is probably the most important feature of the nonlinear theory
\begin{eqnarray*}
m _\nu (u) = \frac{1}{4 \pi} \int _{S ^2} M (u, \theta, \phi)n ^{\nu} d S, \quad \nu =0,1,2,3,
 \end{eqnarray*}
where $n^0=1$, $n^1 =\sin \theta \cos \phi$, $n^2 =\sin \theta \sin \phi$, $n^3=\cos \theta$.
The famous Bondi energy loss formula asserts \cite{Bondi, Sachs}
\begin{eqnarray*}
\frac{d}{du} m _{0} (u)=-\frac{1}{4\pi}\int _{S ^2} (c _{,u}) ^2 +(d _{,u})^2 \leq 0.
\end{eqnarray*}
If $|m(u)|=\sqrt{m _1 ^2 (u)+ m _2 ^2 (u)+m _3 ^2 (u)}\neq 0$, then the Bondi energy-momentum loss formula can be derived \cite{HYZ, Z15}
\begin{eqnarray*}
\frac{d}{du} \Big(m _0 (u) -|m(u)| \Big) = -\frac{1}{4\pi}\int _{S ^2} \Big[(c _{,u}) ^2 +(d _{,u})^2 \Big]
\Big(1- \frac{m_i n^i}{|m|} \Big) \leq 0.
\end{eqnarray*}
Thus the Bondi energy-momentum can be viewed as the total energy-momentum measured after the loss due to the gravitational radiation up to that time.

It was shown, under this natural boundary condition, the Weyl curvature components of Bondi-Sachs metrics satisfy the {\em peeling} property
\begin{eqnarray*}
\Psi_k = O\Big(\frac{1}{r^{5-k}} \Big), \quad k=0, \cdots, 4
\end{eqnarray*}
for zero cosmological constant. This property is crucial for constructing wave forms in numerical simulations.

The wave-like spacetime metrics which radiate energy-momentum are the key feature of gravitational waves. The Bondi-Sachs metrics and the Bondi energy-momentum consist perfectly with the physical picture when the cosmological constant is zero. It is an important question what happens when the cosmological constant is nonzero as cosmological observations actually indicate that the universe has a positive cosmological constant. It has been studied extensively for gravitational waves in this case in recent years, e.g. \cite{GLSWZ, HC, ABK1, ABK2, ABK3, ABK4}. Denote $\Lambda $ the cosmological constant. In \cite{GLSWZ}, a detail asymptotic analysis of Bondi-Sachs metrics for $\Lambda \neq 0$ was provided. By assuming again Sommerfeld's radiation condition, which is natural in numerical simulations, the new boundary condition involving three additional functions $B$, $X$, $Y$ were proposed. In particular, the nonzero Bondi news ensure that $X$, $Y$ must be nonzero. In \cite{HC}, an alternative boundary condition was given in the axi-symmetric case, without assuming Sommerfeld's radiation condition, but deforming 2-sphere with
\[
\gamma=\Lambda f(u, \theta)+\frac{c(u, \theta)}{r}+O\Big(\frac{1}{r^3}\Big)
\]
and taking $B=X=Y=0$. Since the vacuum field equations give $\frac{\partial f }{\partial u} = \frac{c}{3}$, see \cite{HC}, then
\[
f(u, \theta)=f(-\infty, \theta)+\frac{1}{3}\int _{-\infty} ^u c(s, \theta) ds.
\]
Physically, $f(-\infty, \theta)$ exists. Thus, in order that $f(u, \theta)$ exists for any $u<\infty$ and for $u\rightarrow +\infty$, Bondi news must satisfy $\int _{-\infty} ^u  c(s, \theta) ds <\infty$ for any $u<\infty$ and for $u=+\infty$. This boundary condition is rather restricted which actually excludes gravitational waves with $\int _{-\infty} ^u c(s, \theta) ds =\infty$ or $\int _u ^{+\infty} c(s, \theta) ds =\infty$ for some $u<\infty$ or for $u=+\infty$. In a series of papers \cite{ABK1, ABK2, ABK3, ABK4}, asymptotics with $\Lambda >0$ was discussed in framework of conformal compactification, and the linearization theory as well as the quadrupole formula were derived. Some relevant works on the linearization theory can also be found in \cite{B, DH1, DH2}. The papers \cite{S1, S2} discussed the asymptotic vacuum and the electromagnetism Newman-Penrose equations as well as Bondi mass for $\Lambda \neq 0$. The boundary condition in \cite{S1, S2} is essentially equivalent to that given in \cite{HC}.

The peeling property for $\Lambda \neq 0$ was proved previously by Penrose in framework of conformal compactification \cite{Penr},
and by Saw without conformal compactification \cite{S1, S2}. But the boundary condition in \cite{GLSWZ} induces the metric
\[
\frac{e^{2B} \Lambda}{3}  du^2 +2 Y du d\theta +2X du d\phi +d\theta ^2 +\sin^2 \theta d\phi ^2.
\]
on conformal boundary $J^+ $ of Bondi-Sachs metrics. This does not seem to consist with Penrose's framework. In this paper, we show that the peeling property still holds for Bondi-Sachs metrics with nonzero $B$, $X$, $Y$. It is somehow surprising as the Bondi-Sachs metrics do not seem to be (anti-)de Sitter at infinity in current situation. We also construct some nonstationary vacuum Bondi-Sachs metrics without Bondi news, whose Newmann-Penrose quantities fall faster than usual.

The paper is organized as follows:
In Section 2, we introduce the natural boundary condition of Bondi-Sachs metrics with $\Lambda \neq 0$, which is used to derive the peeling property.
In Section 3, we prove the peeling property at infinity for Bondi-Sachs metrics with $\Lambda \neq 0$. The peeling property shows that the cosmological constant affects the experimental data only in a scale of $\Lambda$ which can be ignored at infinity.
In Section 4, we construct some nonstationary vacuum Bondi-Sachs metrics with $B\neq 0$, $\gamma =\delta =0$. This provides some new feature for $\Lambda \neq 0$ that certain gravitational waves may have no news.
In Section 5, we conclude our result and discuss the further work.

\mysection{Natural boundary condition}\ls

Suppose a spacetime has a family of non-intersecting null hypersurfaces given by the level
sets of smooth function $u$. We choose coordinates $x^0 =u$, $x^1=r$, $x^2=\theta$ and $x^3=\phi$
where $r$ is a luminosity distance along the null rays. Following from \cite{GLSWZ}, Bondi-Sachs metrics
for $\Lambda \neq 0$ are also given by
\begin{equation*}
g=-\Big(e^{2\beta}
\frac{V}{r}-r^2h_{\mu \nu}U^\mu U^\nu\Big)du^2-2e^{2\beta}dudr-2r^2h_{\mu \nu}U^\nu du dx^\mu +r^2h_{\mu \nu}dx^\mu dx^\nu
\end{equation*}
where $\mu, \nu =2,3$, $U^2= U$, $U^3= W\csc\theta$,
\[
h\equiv\left(
\begin{array}{ll}
 h_{22}  & h_{23}  \\
 h_{32}  & h_{33}
\end{array}
\right) \equiv\left(
\begin{array}{ll}
 e^{2\gamma}\cosh 2\delta  & \,\,\,\,\,\sinh 2\delta\sin\theta  \\
 \sinh 2\delta\sin\theta  & e^{-2\gamma}\cosh 2\delta\sin^2\theta
\end{array}
\right),
\]
$\beta$, $\gamma$, $\delta$, $V$, $U$, $W$ are functions of $u$, $r$ and points on unit 2-sphere parameterized by $\theta$, $\phi$, i.e.,
they take the same values at $\phi =0$ and $\phi=2\pi$. For $\Lambda >0$, $u$ is only continuous with discontinuous derivatives
across the cosmological horizon, so the above metrics are valid only inside and outside the cosmological horizon.

Now we study the vacuum Einstein field equations with $\Lambda$
\begin{eqnarray}
R_{ij}=\Lambda g_{ij}.  \label{vacuum}
\end{eqnarray}
They reduce to $\Eq (1)$-$\Eq (6)$ of Appendix A in \cite{GLSWZ} together with three supplementary equations
\[
R_{0i}=\Lambda g _{0i}, \quad i =0, 2, 3.
\]
$\Eq (1)$-$\Eq (4)$ are the hypersurface equations which have the form
\[
\begin{aligned}
\beta _{,r} =& \frac{1}{2} r \Big(\gamma ^2 _{,r} \cosh ^2 2 \delta  +\delta _{,r}^2   \Big),\\
\Big(r^4 e^{-2\beta}(e^{2\gamma}U_{,r} \cosh2\delta+W_{,r} \sinh2\delta)\Big)_{,r}=& \Big(\gamma, \delta, \beta \Big)_{, r, \theta, \phi},\\
\Big(r^4 e^{-2\beta}(e^{-2\gamma}W_{,r} \cosh2\delta+U_{,r} \sinh2\delta)\Big)_{,r}=& \Big(\gamma, \delta, \beta \Big)_{, r, \theta, \phi},\\
V_{,r}=&\Lambda r^2 e^{2\beta}+\Big(\gamma, \delta, \beta, U, W \Big)_{, r, \theta, \phi}.
\end{aligned}
\]
$\Eq (5)$-$\Eq (6)$ are the standard equations which have the form
\[
\begin{aligned}
\big(r\gamma\big)_{,ur}\cosh2\delta+2r\Big(\gamma_{,u}\delta_{,r}+\delta_{,u}\gamma_{,r}\Big)\sinh2\delta =&\Big(\gamma, \delta, \beta, U, W \Big)_{, r, \theta, \phi},\\
\big(r\delta \big)_{,ur}-2r\gamma_{,u}\gamma_{,r}\sinh2\delta\cosh2\delta =&\Big(\gamma, \delta, \beta, U, W \Big)_{, r, \theta, \phi}.
\end{aligned}
\]
Three supplementary equations on $R_{02}$, $R_{03}$, $R_{00}$ give
\[
\begin{aligned}
U_{,ur} =&\Big(\big(\gamma, \delta, \beta \big)_{, u, r, \theta, \phi}, \big(U, W, V\big)_{,r, \theta, \phi}, \Lambda \Big),\\
W_{,ur} =&\Big(\big(\gamma, \delta, \beta \big)_{, u, r, \theta, \phi}, \big(U, W, V\big)_{,r, \theta, \phi}, \Lambda \Big),\\
V _{,u} =&\Big(\big(\gamma, \delta, \beta, U, W \big)_{, u, r, \theta, \phi}, V_{,r, \theta, \phi}, \Lambda \Big).
\end{aligned}
\]

Mathematically, we need to solve the above nine equations and recover 4-dimensional spacetime metrics from the deformation of metrics $h$ on 2-dimensional spheres. For given $\gamma$, $\delta$ on certain hypersurface $u=constant$, the four hypersurface equations are used to solve $\beta$ up to an integration function $B(u, \theta, \phi)$, $U$ up to two integration functions $Y(u, \theta, \phi)$, $\frac{_3U(u, \theta, \phi)}{r^3}$, $W$ up to two integration functions $X(u, \theta, \phi)$, $\frac{_3W(u, \theta, \phi)}{r^3}$ and $V$ up to an integration functions $2M(u, \theta, \phi)$.
The standard and supplementary equations evolute $\gamma$, $\delta$, $_3U$, $_3W$ and $M$ to hypersurface $u+\epsilon$ and then the hypersurface equations are used again to solve $\beta$, $U$, $W$ and $V$ on hyperfurface $u+\epsilon$. However, the rigorous proof of the existence is still open, we study only the formal series solutions. Same as \cite{Bondi, Sachs}, we assume Sommerfeld's radiation condition holds, which means that $\gamma$, $\delta$ can be expanded as follows
\[
\begin{aligned}
\gamma&=\frac{c}{r}+\Big(-\frac{1}{6} c^3-\frac{3}{2} d^2 c+C\Big)
   \frac{1}{r^3}+O\Big(\frac{1}{r^4}\Big),\\
\delta&=\frac{d}{r}+\Big(-\frac{1}{6}d^3+\frac{1}{2} c^2d+D\Big)
   \frac{1}{r^3}+O\Big(\frac{1}{r^4}\Big)
\end{aligned}
\]
with the regularity condition $\int _0 ^{2\pi} c(u, \theta, \phi)d\phi =0$ for $\theta =0, \pi$ and for all $u$. For $\Lambda >0$, the above expansions are valid both near and inside the cosmological horizon and at infinity outside the cosmological horizon. Then the hypersurface equations
give
\[
\begin{aligned}
\beta =&B -\frac{c^2+d^2}{4r^2}+O\Big(\frac{1}{r^3}\Big)\\
W=&X+2 e^{2 B} B_{,\phi} \csc \theta\frac{1}{r}+e^{2 B} \Big(2 c
   B_{,\phi}\csc \theta-2 d B_{,\theta}-\bar l \Big) \frac{1}{r^2}+\frac{_3 W}{r^3}+O\Big(\frac{1}{r^4}\Big)\\
U=&Y+2 e^{2 B} B_{,\theta}\frac{1}{r}-e^{2 B} \Big( 2 d B_{,\phi}\csc \theta +2 cB_{,\theta}+l\Big)\frac{1}{r^2}+\frac{_3 U}{r^3}
+O\Big(\frac{1}{r^4}\Big)\\
V=&-\frac{e^{2B}\Lambda}{3}r^3+\Big(\cot\theta Y+\csc \theta X_{,\phi}+Y_{,\theta}\Big)r^2+e^{2 B} \Big(4B_{,\phi}^2 \csc^2\theta\\
  &+2 B_{,\phi \phi} \csc ^2\theta + 2 B_{,\theta} \cot \theta+4 B_{,\theta} ^2 +2 B_{,\theta \theta}+1\Big)r-2 M+O\Big(\frac{1}{r}\Big)
\end{aligned}
\]
where $l = c _{, \theta} +2c \cot \theta +d _{, \phi} \csc \theta$, $\bar l = d_{, \theta} +2d \cot \theta -c _{,\phi} \csc \theta$, and $B$, $X$, $Y$, $_3W$, $_3 U$, $M$ are six arbitrary functions appeared in the integration of four hypersurface equations. They are determinated by initial and boundary conditions. Moreover, $X$, $Y$ satisfy
\[
\begin{aligned}
 \csc \theta X_{,\phi}+\cot \theta Y-Y_{,\theta}=-\frac{2}{3}e^{2B}\Lambda c,\quad
X_{,\theta}-\cot \theta X +\csc\theta Y_{,\phi}=\frac{2}{3}e^{2B}\Lambda d.
\end{aligned}
\]
Let $X=\Lambda a \sint$, $Y=\Lambda b \sint$. Then $a$, $b$ satisfy the following equations
\begin{eqnarray}
a_{,\phi}-\sint b_{,\theta}=-\frac{2}{3} e^{2B} c, \quad b_{,\phi}+\sint a_{,\theta}=\frac{2}{3} e^{2B} d. \label{ab}
\end{eqnarray}

If $\Lambda =0$, that $g_{uu}>0$ requires that $X$, $Y$ must be zero, and $B$ can be transformed to zero by suitable coordinate transformation \cite{Bondi,Sachs}. However, when $\Lambda \neq 0$, the situation is completely different. If $c$, $d$ are nonzero, then $X$, $Y$ must be nonzero
according to the above equations. Moreover, $B$ can not be transformed to zero in general. This raises a new boundary condition. For $\Lambda <0$, the series expansions are taken near $r=\infty$, it does not affect physical properties even if the coefficients involve $\Lambda$. But, for $\Lambda >0$, the series expansions are taken both near and inside the cosmological horizons where $r \sim \sqrt{3 \Lambda ^{-1}}$ is finite, and near $r=\infty$ outside the cosmological horizon. The physical properties are affected if the coefficients of series expansions involve $\Lambda$ near the cosmological horizon.

Formally, we can take derivatives with respect to $\Lambda$ in order to study whether the coefficients of series expansions involve $\Lambda$. From
the two standard equations and three supplementary equations, we know that $B$, $c$, $d$ are free functions which can be chosen not to depend on $\Lambda$. And (\ref{ab}) shows that $a$, $b$ can also be chosen not to depend on $\Lambda$. $_3 U_{,u}$, $_3W_{,u}$ and $M_{,u}$ are determinated by $B$, $a$, $b$ and other coefficients of $\gamma$, $\delta$ which depend on $\Lambda$.

\begin{definition}
The natural boundary condition of Bondi-Sachs metrics with $\Lambda \neq 0$ satisfies (i) Sommerfeld's radiation condition, (ii)$\frac{\partial B }{\partial \Lambda} =\frac{\partial c }{\partial \Lambda}=\frac{\partial d }{\partial \Lambda} =\frac{\partial a }{\partial \Lambda}=\frac{\partial b }{\partial \Lambda} =0$.
\end{definition}

\mysection{Peeling property}\ls

In this section we show the peeling property still holds under the natural boundary condition for $\Lambda \neq 0$.
This is somehow surprising as this boundary condition does not indicate the Bondi-Sachs metrics are de-Sitter or anti-de Sitter
at infinity.

Choose the null tetrad of Bondi-Sachs metrics
\[
\begin{aligned}
  e_0&={\bf l} = e^{-2\beta}\Big(\pu{}-\f V{2r}\pr{} + U\pth{} +\csct W\pf{}\Big) \\
  e_1&={\bf k} =\pr{}  \\
  e_2&={\bf m} = \f{e^{-\gamma}}{r\sqrt{2\cosh{2\delta}}}\Big(1-i\sinh{2\delta}\Big)\pth{}+i\f{e^\gamma \sqrt{\cosh{2\delta}}\csct}{\sqrt{2}r} \pf{}\\
  e_3&={\bar {\bf m}} =\f {e^{-\gamma}}{r\sqrt{2\cosh{2\delta}}}\Big(1+i\sinh{2\delta}\Big)\pth{}-i\f{e^\gamma \sqrt{\cosh{2\delta}}\csct}{\sqrt{2}r} \pf{}
\end{aligned}
\]
with the dual tetrad $\omega^0$, $\omega^1$, $\omega^2$, $\omega^3$. Then the metric is
$$ds^2=-2\omega^0 \omega^1+2\omega^2 \omega^3.$$
The structure coefficients of the tetrad is denoted by $C_{ij}^k$ which satisfy the commutation conditions
$$[e_i,e_j]= C_{ij}^k e_k.$$
By the basic formula in Riemannian geometry
\begin{eqnarray*}
\begin{aligned}
  2\langle\nabla_X Y,Z\rangle =& X\langle Y,Z\rangle+Y\langle Z,X\rangle-Z\langle X,Y\rangle \\
                  & +\langle[X,Y],Z\rangle-\langle[Y,Z],X\rangle+\langle[Z,X],Y\rangle,
\end{aligned}
\end{eqnarray*}
the connection coefficients and the structure coefficients are
$$\Gamma_{ijk}=\f12 \big(C_{ijk}-C_{jki}+C_{kij}\big)$$
where $\Gamma_{ijk}=\Gamma_{ij}^h g_{hk}$, $C_{ijk}=C_{ij}^h g_{hk}$.
The Riemann curvature tensor satisfies
\begin{eqnarray*}
R^i_{jkl} = e_k(\Gamma_{lj}^i)+\Gamma_{lj}^h\Gamma_{kh}^i-C_{kl}^h\Gamma_{hj}^i -e_l(\Gamma_{kj}^i)-\Gamma_{kj}^h\Gamma_{lh}^i.
\end{eqnarray*}
The vacuum Einstein field equations show that the Weyl tensor $W_{ijkl}$ satisfies
\begin{eqnarray*}
  W_{ijkl} = R_{ijkl}-\f23\Lambda g_{i[k} g_{l]j}.
\end{eqnarray*}
The Newmann-Penrose quantities $\Psi _k$, $k=0,\dots,4$, are defined by
\begin{eqnarray*}\label{Psi}
\begin{aligned}
  \Psi_0&= W_{1212}= R_{1212}, \notag\\
  \Psi_1&= W_{1012}= R_{1012}, \notag \\
  \Psi_2&= W_{1230}= R_{1230}-\f{\Lambda}3,\\
  \Psi_3&= W_{0103}= R_{0103},\notag\\
  \Psi_4&= W_{0303}= R_{0303}. \notag
\end{aligned}
\end{eqnarray*}

\begin{theorem} (Peeling Property) Under the natural boundary condition for $\Lambda \neq 0$, $\Psi_k$ satisfy
\begin{eqnarray*}
\Psi_k = -\Big[\big(\Psi_{k} ^{5-k}\big)^0 +O\big(\Lambda \big)\Big]\,\frac{1}{r^{5-k}}+O\Big(\frac{1}{r^{6-k}} \Big), \quad k=0, \cdots, 4,
\end{eqnarray*}
as $r \rightarrow \infty$. Where $\big(\Psi_{4} ^{1}\big)^0$, $\big(\Psi_{3} ^{2}\big)^0$, $\big(\Psi_{2} ^{3}\big)^0$ are determinated by $B$, $a$, $b$. $\big(\Psi_{1} ^{4}\big)^0$ is determinated by $B$, $a$, $b$ and $\Lambda$-independent parts of $_3 W$, $_3 U$. $\big(\Psi_{0} ^{5}\big)^0$
is determinated by $\Lambda$-independent parts of $C$, $D$. Other coefficients are determinated by the coefficients in the series expansions of $\gamma$, $\delta$ and $B$, $a$, $b$, $M$, $_3W$ and $_3U$.
\end{theorem}

\prf The peeling property can be proved by a direct computation. At the beginning, $\Psi_k$ has the unexpected asymptotics
\begin{eqnarray*}
\Psi_k = -\big(\Psi_{k} ^{4-k}\big) \,\frac{1}{r^{4-k}}-\big(\Psi_{k} ^{5-k}\big)\,\frac{1}{r^{5-k}}+O\Big(\frac{1}{r^{6-k}} \Big),
\quad k=0, \cdots,4,
\end{eqnarray*}
where coefficients are given by $B$, $X$, $Y$, $c$, $d$ and other $\Lambda$-independent functions appeared in the series expansions of $\gamma$, $\delta$. It is mysterious that $\big(\Psi_{k} ^{4-k}\big)=0$ when $c$, $d$ are substituted in terms of (\ref{ab}). The precise terms of $\big(\Psi_{k} ^{5-k}\big)^0 $ are given in the appendix. \qed

The peeling property shows that the cosmological constant affects the experimental data only in a scale of $\Lambda$ which can be ignored at infinity. However, near the cosmological horizon when $\Lambda>0$, the asymptotic behaviors of these coefficients become sophisticated. But we still conjecture that $\frac{\big(\Psi_{4} ^{1}\big)^0 }{r}$ is the slowest fall-off term.

\mysection{Gravitational waves without Bondi news}\ls

As it is not known how to define the Bondi energy-momentum equipped with the energy-momentum loss property, it may not be suitable to regard gravitational waves as the Bondi-Sachs metrics which radiate energy when $\Lambda \neq 0$. Instead, we may think them as nonstationary vacuum Bondi-Sachs metrics.

Note that $\big(\Psi_{4} ^{1}\big)^0 $ may be nonzero and the Bondi-Sachs metrics may be nonstationary when $B$ are nontrivial and $a=a(u)$, $b=b(u)$ are functions of $u$ only. It implies $c=d=0$ and indicates that there exist gravitational waves without Bondi news. This is the new feature only for $\Lambda \neq 0$ and they may be referred as $B$-gravitational waves. In this section we construct some nonstationary vacuum Bondi-Sachs metrics with $B\neq 0$, $\gamma =\delta =0$.

Similar to \cite{GLSWZ}, we take the axi-symmetric Bondi-Sachs metrics with $X=0$, $Y=\sin \theta  \sigma(u)$, $M=\tau(u) e^{-4B}$. Then the metrics
\begin{eqnarray}
\begin{aligned}
g=&-e^{2B} \Big[-\frac{\Lambda}{3} e^{2 B} r^2+2 \cos \theta  \sigma r
+e^{2 B }\Big(4 B_{,\theta} ^2+2 \cot \theta  B_{,\theta}+2 B_{,\theta \theta} +1\Big) -\frac{2M}{r}\Big]du^2 \\
&-2e^{2B}dudr +r^2 \Big[ \Big(\sin \theta  \sigma +\frac{2 e^{2 B } B_{,\theta} }{r}\Big)du-d\theta \Big]^2
+r^2 \sin^2\theta d\phi ^2 \label{axi}
\end{aligned}
\end{eqnarray}
are vacuum if and only if $B(u,\theta)$ satisfies
\begin{eqnarray}
\begin{aligned}
&e^{4 B}
   \Big[16 \cot \theta  B_{,\theta} ^3+4 B_{,\theta \theta}^2-4 B_{,\theta} ^2
   \Big(\cot ^2\theta -4 B_{,\theta \theta}\Big)-B_{,\theta \theta}\cot^2 \theta\\
   &+2 \cot \theta  B_{,\theta \theta \theta}+B_{,\theta}  \Big(\cot \theta  \big(\csc ^2\theta
   +16 B_{,\theta \theta}+2\big)+8 B_{,\theta \theta \theta}\Big)+B_{,\theta \theta \theta \theta}\Big]\\
&+4 M B_{,u}-2 M_{,u}-6 \cos \theta  M \sigma-3\sin \theta M_{,\theta} \sigma =0. \label{B}
\end{aligned}
\end{eqnarray}

For any function $C(u)>1$, we choose
\begin{eqnarray*}
B=\frac{\ln(C(u)+\cos\theta)}{2},\quad M=\frac{\tau(u)}{(C(u)+\cos\theta)^2}.
\end{eqnarray*}
It is easy to check that this $B$ satisfies the following equation
\begin{eqnarray*}
\begin{aligned}
&16 \cot \theta  B_{,\theta} ^3+4 B_{,\theta \theta}^2-4 B_{,\theta} ^2
   \big(\cot ^2\theta -4 B_{,\theta \theta}\big)-B_{,\theta \theta}\cot^2 \theta\\
   &+2 \cot \theta  B_{,\theta \theta \theta}+B_{,\theta}  \big(\cot \theta  (\csc ^2\theta
   +16 B_{,\theta \theta}+2)+8 B_{,\theta \theta \theta}\big)+B_{,\theta \theta \theta \theta}=0.
\end{aligned}
\end{eqnarray*}
Now we derive $\sigma $ so that equation (\ref{B}) holds. For the above $B$ and $M$,
\begin{eqnarray*}
B_u =\frac{C'}{2(C+\cos\theta )}, \, M_u =\frac{\tau '}{(C+\cos\theta )^2}-\frac{2\tau C'}{(C+\cos\theta )^3}, \, M_\theta =\frac{2 \tau \sin \theta}{(C+\cos\theta )^3}.
\end{eqnarray*}
Thus (\ref{B}) reduces to
\begin{eqnarray*}
\frac{2\tau C'}{(C+\cos\theta)^3}-\frac{2\tau'}{(C+\cos\theta)^2} +\frac{4\tau C'}{(C+\cos\theta)^3}
-\frac{6\tau \sigma \cos\theta}{(C+\cos\theta)^2}-\frac{6\tau \sigma \sin ^2 \theta}{(C+\cos\theta)^3}=0.
\end{eqnarray*}
Therefore
\begin{eqnarray*}
3\tau C' -\tau ' C -3 \tau \sigma - \big(\tau ' +3 \tau \sigma C\big)\cost=0.
\end{eqnarray*}
As $\tau$, $\sigma$, $C$ are functions of $u$, it gives
$$3\tau C' -\tau ' C -3 \tau \sigma =0, \qquad \tau ' +3 \tau \sigma C=0.$$
Thus,
\begin{eqnarray*}
\sigma =-\frac{C'}{C^2 -1 }, \qquad \tau=m \big(C^2 -1 \big)^\frac{3}{2}
\end{eqnarray*}
where $m$ is constant.

Substituting them into (\ref{axi}), we obtain the following exact vacuum Bondi-Sachs metrics without Bondi news
\begin{eqnarray*}
\begin{aligned}
g=&-\Big[-(C+\cos \theta)^2  \frac{\Lambda}{3}r^2  - \sin ^2\theta\Big(\frac{C'}{C^2 -1}\Big)^2 r^2
-2(C\cos\theta +1)\frac{ C'}{C^2 -1 }r \\
&+C^2 -1 -\frac{2m(C^2 -1)^\frac{3}{2}}{r(C+\cos \theta)}\Big]du^2-2 \big(C+\cos \theta \big)du dr\\
&+2 r \sin \theta \Big(1+\frac{C'}{C^2 -1}r\Big)du d\theta+r^2\Big(d\theta^2 +\sin^2\theta d\phi^2 \Big).
\end{aligned}
\end{eqnarray*}
The above metrics have black holes when $m \neq 0$. Moreover, the Newmann-Penrose quantity $\Psi_k$ satisfy
\begin{eqnarray*}
\begin{aligned}
\Psi_0&=\Psi_1=0,\qquad \Psi_2=-\frac{m(C^2-1)^{\frac{3}{2}}}{r^3(C+\cos\theta)^3},\\
\Psi_3&=\frac{3m\sin\theta(C^2-1)^{\frac{3}{2}}}{\sqrt{2} r^3(C+\cos\theta)^4},\quad
\Psi_4=-\frac{3m\sin^2\theta(C^2-1)^{\frac{3}{2}}}{r^3(C+\cos\theta)^5}.
\end{aligned}
\end{eqnarray*}
They fall faster than usual.

\mysection{Conclusion and future work}\ls

We conclude that, for Bondi-Sachs metrics with $\Lambda \neq 0$, Sommerfeld's radiation condition together with nontrivial $\Lambda$-independent
functions $B$, $c$, $d$, $a$, $b$ provide a boundary condition. This boundary condition is natural in three aspects that it consists with nontrivial Bondi news, gives rise to the peeling property and features new $B$-gravitational waves without Bondi news. Moreover, we observe that, under the natural boundary condition, the cosmological constant affects the experimental data only in a scale of $\Lambda$ which can be ignored. We also notice that there exist $B$-gravitational waves, which Newman-Penrose quantities fall faster than usual. These $B$-gravitational waves may be missed in the experimental data. In the future, we shall study the analogue of the Bondi energy-momentum for $\Lambda \neq 0$.

\bigskip

{\small {\it Acknowledgement.}
This work is partially supported by the National Science Foundation of China (grants 11571345, 11731001) and HLM, NCMIS, CEMS, HCMS of Chinese Academy of Sciences.}

\bigskip
\begin{center}
Appendix A: Formulas for $\big(\Psi_{k} ^{5-k}\big)^0 $
\end{center}

Denote $[_3U]$, $[_3W]$, $[C]$ and $[D]$ the $\Lambda$-independent parts of $_3U$, $_3W$, $C$ and $D$ respectively. Lower indices with comma omitted represent partial derivatives.
\begin{align*}
(\Psi_4^1)^0=&e^{-6 B} \Big(-\f32 a_{uu\phi}+\f32 b_{uu\phi} \sin \theta+
9 a_{u \phi} B_u-12 a_{\phi } B_u^2+3 a_{\phi }B_{uu}\\
&+12 b_{\theta } B_u^2 \sin \theta-9 b_{u\theta} B_u \sin \theta-3 b_{\theta } B_{uu} \sin \theta
+4 e^{4 B} B_{\theta } B_{u\theta}\\
&-e^{4 B} B_{u\theta} \cot \theta+e^{4 B} B_{u \theta \theta} -4e^{4 B} B_{u \phi} B_{\phi } \csc ^2\theta-e^{4 B} B_{u \phi \phi}\csc ^2\theta \Big)\\
&+i e^{-6 B} \Big(-\f32 a_{uu\phi} \sin\theta-\f32 b_{uu\phi}
-12 a_{\theta } B_u^2 \sin \theta+9 a_{u\theta} B_u
   \sin \theta\\
&+3 a_{\theta } B_{uu} \sin \theta+9 b_{u \phi} B_u-12 b_{\phi } B_u^2
+3 b_{\phi } B_{uu}-4 e^{4 B} B_{u\theta} B_{\phi } \csc \theta \\
&-2 e^{4 B}B_{u \phi} \csc \theta-4 e^{4 B} B_{\theta } B_{u \phi} \csc
   \theta+2 e^{4 B} B_{u \phi} \cot \theta \csc \theta\Big)\\\\
  (\Psi_3^2)^0=&\f{e^{-4B}}{2\sqrt{2}}\Big(B_{\theta } (-24 a_{\phi } B_u+12 a_{u\phi}+24 b_{\theta } B_u \sin \theta-12 b_{u\theta} \sin \theta\\
&-2 e^{4 B} \csc ^2\theta+8 e^{4 B}B_{\theta \theta }+4 e^{4 B})+24 a_{\theta } B_u B_{\phi }+12 a_{\phi } B_u \cot \theta\\
&+6 a_{\phi } B_{u\theta}-12 a_{u\theta} B_{\phi }-6a_{\theta } B_{u\phi}-6 a_{u\phi} \cot \theta-6 b_{\theta \theta } B_u \sin \theta\\
&+24 b_{\phi } B_u B_{\phi } \csc \theta-6 b_{\phi \phi } B_u\csc \theta+6 b B_u \sin \theta-18 b_{\theta } B_u \cos \theta\\
&-6 b B_u \csc \theta+6 b B_u \cos \theta \cot \theta-6 b_{\theta } B_{u\theta}\sin \theta-6 b_{\phi } B_{u\phi} \csc \theta\\
&-12 b_{u\phi} B_{\phi } \csc \theta
   -3 b_u \sin \theta+3 b_u \csc \theta-3 b_u \cos \theta\cot \theta+9 b_{u\theta} \cos \theta\\
&+3 b_{ u\theta\theta} \sin \theta+3 b_{ u\phi\phi} \csc \theta+2 e^{4 B} B_{\theta \theta }\cot \theta+2 e^{4 B} B_{\theta \phi \phi } \csc ^2\theta\\
\end{align*}
\begin{align*}
&-4 e^{4 B} B_{\phi \phi } \csc ^2\theta (\cot \theta
   -2 B_{\theta })+8 e^{4 B} B_{\theta }^2\cot \theta+2 e^{4 B} B_{\theta \theta \theta }\Big)\\
&+\f{i e^{-4B}}{2\sqrt{2}}\Big(-4 B_{\phi } (6 a_{\phi } B_u \csc \theta-3 a_{u\phi} \csc \theta-6 b_{\theta } B_u
   +3 b_{u\theta}\\
&+2 e^{4 B} B_{\theta \theta } \csc \theta+2e^{4 B} B_{\phi \phi } \csc ^3\theta+e^{4 B} \csc \theta+2 e^{4 B} B_{\theta } \cot \theta \csc \theta)\\
&+6 a_{\theta \theta } B_u \sin \theta+6 a_{\phi\phi } B_u \csc \theta-6 a B_u \sin \theta-24 a_{\theta } B_{\theta } B_u \sin \theta\\
&+18 a_{\theta } B_u \cos \theta+6 a B_u \csc \theta-6 a B_u \cos \theta \cot \theta+6 a_{\theta } B_{u\theta} \sin \theta\\
&+12 a_{u\theta} B_{\theta } \sin \theta
   +6 a_{\phi } B_{u\phi} \csc \theta+3 a_u\sin \theta-3 a_u \csc \theta\\
&+3 a_u \cos \theta \cot \theta-9 a_{u\theta} \cos \theta-3 a_{ u\theta\theta} \sin \theta
   -3a_{ u\phi\phi} \csc \theta-24 b_{\phi } B_{\theta } B_u\\
&+12 b_{\phi } B_u \cot \theta+6 b_{\phi } B_{u\theta}-6 b_{\theta } B_{u\phi}
   +12 b_{u\phi} B_{\theta }-6 b_{u\phi} \cot \theta\\
&-2 e^{4 B} B_{\theta \theta \phi } \csc \theta-2 e^{4 B} B_{\theta \phi } \cot \theta \csc \theta
   -2 e^{4 B} B_{\phi \phi \phi } \csc ^3\theta\Big),\\\\
(\Psi_2^3)^0=&\f{e^{-6B}}{4}\Big(4 e^{4 B} M+3 (12 a_{\phi } b_{\theta } B_u \sin \theta-12 a_{\theta } b_{\phi } B_u \sin \theta-3 a_{\phi } b_{u\theta} \sin \theta\\
&+3a_{u\theta} \sin \theta (a_{\theta } \sin \theta+b_{\phi })+3 a_{u\phi} (a_{\phi }-b_{\theta } \sin \theta)+3 a_{\theta }b_{u\phi} \sin \theta\\
&-4 e^{4 B} a_{\phi \phi } B_{\phi } \csc ^2\theta-8 e^{4 B} a_{\phi } B_{\theta }^2
+16 e^{4 B} a_{\theta } B_{\theta } B_{\phi }+8 e^{4B} a_{\phi } B_{\theta } \cot \theta\\
&-12 e^{4 B} a_{\theta } B_{\phi } \cot \theta+8 e^{4 B} a_{\phi } B_{\phi }^2 \csc ^2\theta
-4 e^{4 B} a_{\theta \theta }
   B_{\phi }-6 a_{\theta }^2 B_u \sin ^2\theta\\
&-6 a_{\phi }^2 B_u-4 e^{4 B} b_{\theta \theta } B_{\theta } \sin \theta
-8 e^{4 B} b_{\theta } B_{\phi }^2 \csc \theta+16
   e^{4 B} b_{\phi } B_{\theta } B_{\phi } \csc \theta\\
&-8 e^{4 B} b_{\phi } B_{\phi } \cot \theta \csc \theta-4 e^{4 B} b_{\phi \phi } B_{\theta } \csc \theta+8
   e^{4 B} b_{\theta } B_{\theta }^2 \sin \theta\\
&+4 b e^{4 B} B_{\theta } \sin \theta-12 e^{4 B} b_{\theta } B_{\theta } \cos \theta+4 b e^{4 B} B_{\theta } \cos \theta \cot \theta\\
&-4 b e^{4 B} B_{\theta } \csc\theta-6 b_{\theta }^2 B_u \sin ^2\theta-6 b_{\phi }^2 B_u+3 b_{\theta } b_{u\theta} \sin ^2\theta+3 b_{u\phi} b_{\phi })\Big)\\
&+\f{3i e^{-6B}}{4}\Big(a_{\theta } \sin \theta (-3 a_{u\phi}+3 b_{u\theta} \sin \theta
   +e^{4 B} (-4 B_{\phi \phi } \csc ^2\theta+8 B_{\theta } \cot \theta\\
&+4B_{\theta \theta }-3 \cot ^2\theta+3))-3 a_{u\theta} b_{\theta } \sin ^2\theta+3 a_{\phi } b_{u\phi}
   -3 a_{u\phi} b_{\phi}\\
&+e^{4 B} a_{\theta \theta } (4 B_{\theta } \sin \theta-5 \cos \theta)-e^{4 B} a_{\theta \theta \theta } \sin \theta+8 e^{4 B} a_{\phi } B_{\theta \phi} \csc \theta\\
&-e^{4 B} a_{\theta \phi \phi } \csc \theta+4 e^{4 B} a_{\phi \phi } B_{\theta } \csc \theta-2 e^{4 B} a_{\phi \phi } \cot \theta \csc \theta\\
&-4 ae^{4 B} B_{\theta } \sin \theta+3 a e^{4 B} \cos \theta +4 a e^{4 B} B_{\theta } \csc \theta+3 a e^{4 B} \cos \theta \cot ^2\theta\\
&-4 a e^{4 B} B_{\theta }\cos \theta \cot \theta-3 a e^{4 B} \cot \theta \csc \theta+3 a_{u\theta} a_{\phi } \sin \theta-8 e^{4 B} b_{\theta } B_{\theta \phi }\\
&+e^{4 B}b_{\theta \phi } \cot \theta
-4 e^{4 B} b_{\phi \phi } B_{\phi } \csc ^2\theta-4 e^{4 B} b_{\phi } B_{\phi \phi } \csc ^2\theta
   -3 e^{4 B} b_{\phi } \cot ^2\theta\\
&+4 b e^{4 B} B_{\phi } \cot ^2\theta+4 e^{4 B} b_{\phi } B_{\theta } \cot \theta-4 e^{4 B} b_{\theta } B_{\phi } \cot \theta +3 e^{4 B} b_{\phi } \csc ^2\theta\\
&-4 b e^{4 B} B_{\phi } \csc ^2\theta+e^{4 B} b_{\phi \phi \phi } \csc ^2\theta+4 e^{4 B} b_{\phi } B_{\theta \theta }-4 e^{4 B} b_{\theta \theta } B_{\phi }\\
&+e^{4 B}b_{\theta \phi \phi }-e^{4 B} b_{\phi }+4 b e^{4 B} B_{\phi }+3 b_{u\theta} b_{\phi } \sin \theta-3 b_{\theta } b_{u\phi} \sin \theta\Big),\\\\
\end{align*}
\begin{align*}
  (\Psi_1^4)^0=&\f{3e^{-4B}\csc\theta}{32\sqrt{2}}\Big(-16 e^{2 B} \sin \theta [_3U]
     +36 a_{\theta } \sin \theta (3 a_{\theta \theta } \sin ^2\theta+2 a_{\phi \phi }\\
&-10 b_{\phi } B_{\theta } \sin\theta+b_{\theta \phi } \sin \theta+11 b_{\phi } \cos \theta)
     +180 a_{\phi } b_{\theta } B_{\theta }\\
&-180 a_{\phi } b_{\theta } B_{\theta } \cos 2\theta
+54a_{\phi } b_{\theta \theta } \cos 2\theta-54 a_{\theta \theta } b_{\phi } \cos 2\theta\\
&-36 a_{\theta \phi } \sin \theta (b_{\theta } \sin \theta-a_{\phi})
-198 a_{\phi } b_{\theta } \sin 2\theta-54 a_{\phi } b_{\theta \theta }\\
&+54 a_{\theta \theta } b_{\phi }+72 a_{\phi \phi } b_{\phi }-72 a_{\phi } b_{\phi \phi}-180 a_{\phi }^2 B_{\theta } \sin \theta+144 a_{\phi }^2 \cos \theta\\
&+144 b_{\phi }^2 \cos \theta-36 a_{\theta }^2 \sin ^2\theta (5 B_{\theta } \sin \theta-7 \cos \theta)
   -180b_{\phi }^2 B_{\theta } \sin \theta\\
&-135 b_{\theta }^2 B_{\theta } \sin \theta+45 b_{\theta }^2 B_{\theta } \sin 2\theta
   +81 b_{\theta } b_{\theta \theta } \sin\theta-27 b_{\theta } b_{\theta \theta } \sin 2\theta\\
&+36 b_{\theta \phi } b_{\phi } \sin \theta+72 b_{\theta } b_{\phi \phi } \sin\theta
   +63 b_{\theta }^2 \cos \theta-63 b_{\theta }^2 \cos 2\theta\Big)\\
&+\f{3i e^{-4B}\csc\theta}{16\sqrt{2}}\Big(-8 e^{2 B} \sin \theta [_3W]
    +18 a_{\theta } \sin \theta (a_{\theta \phi } \sin \theta+2 a_{\phi } \cos \theta\\
&-10 b_{\phi } B_{\phi}-b_{\theta \theta } \cos 2\theta+b_{\theta \theta }+3 b_{\phi \phi })
   +180 a_{\phi } b_{\theta } B_{\phi } \sin \theta\\
&+36 a_{\theta \theta } \sin ^2\theta(a_{\phi }-b_{\theta } \sin \theta)-18 a_{\phi } b_{\theta \phi } \sin \theta+18 a_{\theta \phi } b_{\phi } \sin \theta\\
&-54 a_{\phi \phi } b_{\theta }\sin \theta-90 a_{\theta }^2 B_{\phi } \sin ^2\theta
-90 a_{\phi }^2 B_{\phi }+54 a_{\phi } a_{\phi \phi }-45 b_{\theta }^2 B_{\phi }\\
&+45 b_{\theta }^2 B_{\phi } \cos2\theta
   -90 b_{\phi }^2 B_{\phi }-18 b_{\theta \theta } b_{\phi } \cos 2\theta+9 b_{\theta } b_{\theta \phi }\\
&-9 b_{\theta } b_{\theta \phi } \cos 2\theta+18b_{\theta } b_{\phi } \sin 2\theta+18 b_{\theta \theta } b_{\phi }+54 b_{\phi } b_{\phi \phi }\Big),\\\\
\qquad \quad (\Psi_0^5)^0=&6\Big([C]+i[D]\Big).\\\\
\end{align*}

\end{document}